%% file: main.tex
\documentclass{article}

\usepackage{arxiv}

\usepackage[utf8]{inputenc} % allow utf-8 input
\usepackage[T1]{fontenc}    % use 8-bit T1 fonts
\usepackage{url}            % simple URL typesetting
\usepackage{booktabs}       % professional-quality tables
\usepackage{amsfonts}       % blackboard math symbols
\usepackage{nicefrac}       % compact symbols for 1/2, etc.
\usepackage{microtype}      % microtypography
\usepackage{lipsum}
\usepackage{graphicx}
\usepackage{orcidlink}
\usepackage{comment}
\definecolor{VUB_blauw}{rgb}{0.1529, 0.2667, 0.5529}
\usepackage{hyperref}       % hyperlinks
\hypersetup{
    colorlinks,%
    citecolor=VUB_blauw,%
    filecolor=VUB_blauw,%
    linkcolor=VUB_blauw,%
    urlcolor=VUB_blauw
}
\graphicspath{ {./images/} }

\usepackage{xcolor}
\newcommand{\red}[1]{\textcolor{black}{#1}}

\usepackage{amsmath} 
\usepackage{pdfpages}

\AddToHook{shipout/background}{%
  \ifnum\value{page}=1 % Check if the current page is 2
    \pagenumbering{Roman} % Change numbering to Roman numerals
    \setcounter{page}{1} % Set the page number to II
  \fi
  \ifnum\value{page}=2 % Check if the current page is 2
  \pagenumbering{arabic} % Revert to Arabic numerals
  \setcounter{page}{2} % Set the page number to 3
  \fi
}

\title{Turning Citation Networks Inside Out: Studying Science Using Content-Based Knowledge Graphs from LLM-Derived Taxonomies}
\runningtitle{Turning Citation Networks Inside Out}

\author{
  Seorin Kim\textsuperscript{1,*} \\ 
  \orcidlinkc{0009-0001-5086-2704} \\
  \And
  Vincent Holst \textsuperscript{1} \\
  \orcidlinkc{0009-0002-4117-4966} \\
  \And
  Vincent Ginis \textsuperscript{1,2} \\ 
  \orcidlinkc{0000-0003-0063-9608} \\
  \and
  \textsuperscript{1}Data Analytics Lab, Vrije Universiteit Brussel, 1050, Brussel, Belgium \\ 
  \textsuperscript{2}School of Engineering and Applied Sciences, Harvard University, Cambridge, Massachusetts, 02138, USA
}
    
\begin{document}
\maketitle
\renewcommand{\thefootnote}{} % Suppress footnote number
% \footnotetext{\includegraphics[height=1em]{panda2.png}\, Corresponding author: seorin.kim@vub.be}
\footnotetext[1]{*\,
Corresponding author: \href{mailto:seorin.kim@vub.be}{seorin.kim@vub.be}}
\renewcommand{\thefootnote}{\arabic{footnote}} % Restore footnote numbering (optional)
\thispagestyle{plain} 

\begin{abstract}
Scientific fields are often mapped using citations and metadata, despite knowledge being transmitted primarily through content. We introduce an “inside-out” approach that reconstructs field structure directly from text by representing each paper as a small set of interpretable knowledge components. Using a large language model to induce domain-specific taxonomies and label papers, each publication is encoded as a triplet of measure, data type, and research-question type. These triplets define a knowledge graph with edges weighted by shared papers. Applied to 617 studies on intergenerational wealth mobility, the graph reveals a stable methodological backbone centered on regression-based mobility measures, alongside substantial temporal variation in component recombination. We further utilize normalized betweenness-to-connectivity ratios to identify components and pairings that act as structural bridges disproportionate to their prevalence. This content-derived, taxonomy-driven mapping complements citation-based approaches by exposing the evolving architecture of methods, data, and questions that define a field.
% Scientific fields are usually studied using citations and other metadata, even though papers primarily transmit knowledge through their content. We present an “inside-out” approach that reconstructs a field structure directly from textual content by converting each paper into a small set of interpretable knowledge components. Using a large language model to induce domain-specific taxonomies and to label papers, we represent every publication as a triplet of measure, data type, and research-question type, and build a knowledge graph whose edges are weighted by shared papers. On a tightly curated corpus of 617 studies of intergenerational wealth mobility, this content-level graph reveals a stable methodological backbone—regression-based mobility measures, alongside substantial temporal variation in how dominant components are recombined into pairs and triads. Beyond frequency, we utilize normalized betweenness-to-connectivity ratios that highlight components and pairings that act as disproportionately important structural bridges relative to their prevalence, pointing to combinations that are central to the knowledge network despite limited uptake. Our work demonstrates how content-derived, taxonomy-driven graphs can complement citation-based science mapping by exposing the evolving architecture of methods, data, and questions that define a research field.
\end{abstract}

% keywords can be removed

\section{Teaser}
Large language models can convert paper content into an interpretable knowledge graph that reveals stable methodological backbones and overlooked bridges in a research field.
% Academic papers are written with the purpose of conveying knowledge or ideas discovered by an individual or a group of individuals. While the main findings of a paper may later lose their validity within a field, its legacy, including its methodology, the type of data it employed, or the research question it addressed, may continue to propagate across scientific domains. In this sense, an academic paper is a multifaceted object whose value lies in its content, and the network of science must be understood accordingly. Grounded in this perspective, we bring knowledge exploration to the content level by leveraging a large language model. Using a small yet highly curated dataset, we demonstrate how this approach provides a deeper view of how knowledge components interact to form a scientific field.

\section{Introduction}
The evolution and organization of science are most commonly studied through citation networks, which use reference relations between academic articles as proxies for the flow of scientific knowledge. Here, we take a radically different approach. Rather than treating papers as the central units of analysis, we deliberately move away from document-centered representations altogether. We propose a content-based knowledge graph in which scientific structure is reconstructed directly from conceptual categories inferred from the text, without relying on citations or other bibliometric metadata. In this sense, the framework turns the usual science-of-science perspective \textit{inside-out}: instead of studying science via the relations between papers, we study how the substantive components of research assemble into a coherent field.

To operationalize this approach at scale, we leverage recent advances in large language models (LLMs) for taxonomy generation and automatic annotation. Prior work has demonstrated the capacity of LLMs to extract structured information from scientific \cite{seltmann2025leveraging, reason2024artificial, lu2025karma} texts as well as to automate categorization tasks \cite{lan2025nlp, wu2025automated}. Here, we use LLMs to generate and assign a small number of interpretable categories to each publication. The resulting graph is constructed solely from these categories and their co-occurrences. By restricting the representation to three node types, each paper is effectively mapped to a triplet of conceptual features, yielding a compact yet expressive graph that allows both static and temporal structural analysis. 

We apply this framework to a highly curated corpus of 617 publications on intergenerational wealth mobility. The corpus includes empirical studies that employ wealth mobility measures as well as theoretical contributions analyzing these measures. Accordingly, the three conceptual dimensions -- \textit{Measures}, \textit{Data types}, and \textit{Research-Question Types} -- are chosen to reflect core methodological and substantive choices in this literature. We analyze the resulting graph using a set of exploratory structural measures, including node, pair, and triangle counts, as well as degree, strength (weighted degree), and node- and edge-level betweenness. In addition, we examine ratios of betweenness to degree (and to count) to identify components that occupy structurally central positions despite limited overall prevalence. 

Applying these measures reveals that regression-based measures have formed a persistent methodological backbone of the wealth mobility literature since the 2006-2010 period, appearing among the most frequent pairs and triangles. Moreover, comparing node- and edge-level betweenness ratios shows that structural variability arises primarily in how components are combined (pair-wise dynamics) than in which components dominate (node-wise dynamics). In other words, while popular nodes are persistent over periods, their combinations vary over periods. 

Through the case study, we show that taking this \textit{inside-out} approach, namely focusing on the content of the academic articles rather than their citation relations, can recover meaningful signals in a controlled setting.

\subsection{\red{Previously Explored Methods}}
Understanding the structure and evolution of scientific knowledge requires analytical tools that can capture the conceptual, methodological, and epistemological configurations that underlie a field. Here, the advent of large bibliometric datasets, built on seminal works on citation indexing \cite{garfield1955citation} and networks of scientific papers \cite{ price1965networks}, has enabled data-driven approaches based, for instance, on \red{citation analysis \cite{radicchi2011citation} and keyword co-occurrences \cite{radhakrishnan2017novel}}. However, these techniques rely on the implicit assumption that citations and keywords are reliable proxies for publication content and academic influence \cite{liu2023scientometrics, borner2011introduction}. 

This limitation becomes particularly evident when citation-based representations are used to study knowledge recombination and novelty, as in e.g., \cite{uzzi2013atypical, WANG20171416}. In \cite{FONTANA2020newandatypical}, the authors show that the aforementioned two indicators are largely driven by interdisciplinarity and structural properties of citation \red{networks.} This points to a broader difficulty of citation analysis, already emphasized \red{in} classic work on the heterogeneous and often rhetorical motives for citing prior work \cite{nigel1977referencing, brooks1986evidence, cozzens1989citations, macroberts1989problems}. Indeed, further studies indicate that many citations misrepresent or only weakly engage with the actual content of the cited work, or are merely incidental mentions \cite{greenberg2009citation, horbach2021meta, moravcsik1975some, valenzuela2015identifying}. Taken together, these observations suggest that treating citations as proxies for content risks producing a distorted view of the underlying knowledge space.

Keywords co-occurrence analyses similarly aim to identify the knowledge structure of a field, but by linking the keywords of publications, which seem to be associated with the research content more than citations do, and weighting edges by their co-occurrence frequencies \cite{radhakrishnan2017novel}. Accordingly, this approach is commonly used in systematic reviews \cite{grames2019automated,yuan2022trends}. Despite the fact that keywords provide a semantic representation of publications, the incoherent consistency and quality affect the reliability of the analyses \cite{sampagnaro2023keyword}. 

More recently, natural language processing techniques such as word embeddings and concept-extraction models have emerged to capture semantic relationships in scientific texts, thereby providing a more content-level representation of research. In particular, leveraging its high-dimensional semantic structure, word embedding approaches have been used to quantify novelty by measuring semantic distances or detecting shifts in conceptual neighborhoods over time \cite{aceves2024mobilizing,yin2023wordembedding,shibayama2021measuring}. However, this high dimensionality is a limitation of this approach, as it lacks interpretability and therefore, the network structure is not inherently meaningful \cite{aceves2024mobilizing}. Moreover, it is sensitive to training data and parametrization \cite{yin2023wordembedding}, making it difficult to link specific dimensions to epistemic constructs or methodological lineages. 

Complementing this line of work, the concept-extraction approach by \cite{tosi2021scikgraph, tosi2022understanding} identifies concepts directly from full-texts using BabelNet and represents them as a concept co-occurrence knowledge graph, treating concepts as the fundamental units of analysis. This knowledge graph is clustered into overlapping topics and summarized via centrality-based key concepts and key phrases. This framework moves beyond metadata and provides a more interpretable view of research fields as subfields emerge as clusters of meaningfully related concepts. Nevertheless, it relies heavily on Babelfy/BabelNet coverage, hand-tuned pruning threshold, and post-hoc cluster agglomeration. Therefore, it may struggle with highly technical domains that use specialized jargon and symbols, interdisciplinary boundary concepts, and large-scale, fully automated deployment.

\subsection{The Framework} \label{sec:framework}

\begin{figure}
    \centering
    \includegraphics[width=0.8\linewidth]{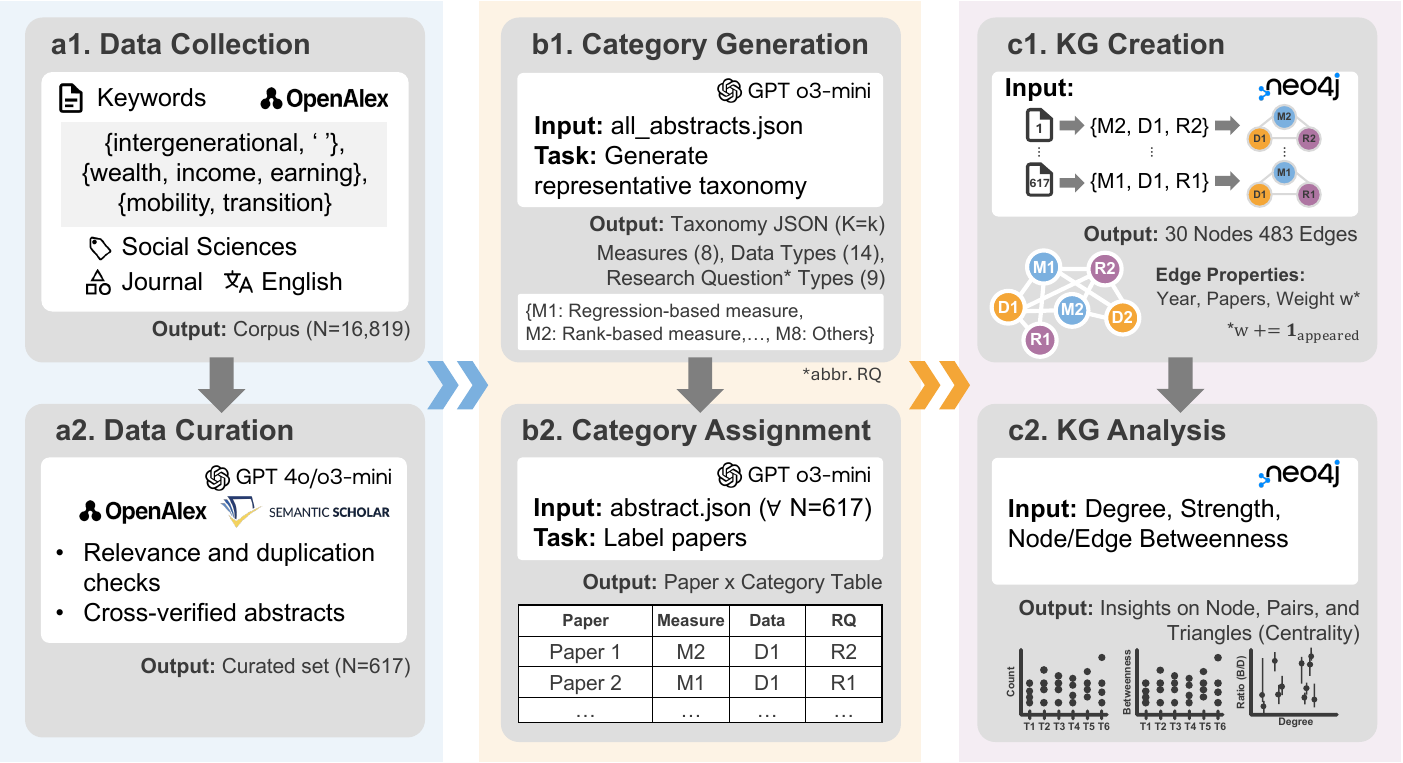}
    \caption{\textbf{Framework overview.} \textbf{a.} From an initial corpus of 16,819 publications, LLM-assisted curation identifies 617 relevant studies on intergenerational mobility. \textbf{b.} Using LLM-driven classification, each paper is assigned to three taxonomies -- measures, data types, and research-question types -- derived from the LLM-generated category sets. \textbf{c.} These taxonomies form the basis of a knowledge graph where edges represent co-occurrence between categories, weighted by the number of shared papers, and the knowledge graph is analyzed with centrality measures.}
    \label{fig:overview}
\end{figure}

We consider a literature review on intergenerational wealth mobility research as a case study in order to test our content-based co-occurrence knowledge graph approach to exploring a scientific field. For this literature review, we aimed to investigate how different measures are used to study wealth mobility over time. We therefore sought papers that either empirically study wealth mobility using data or theoretically discuss its measurement. 

Because similar statistical models are often employed to study income and earnings mobility, these terms were also included in the search process. As shown in Fig.~\ref{fig:overview}, an initial pool of 16,819 publications was retrieved from OpenAlex \cite{priem2022openalex} and subsequently filtered using OpenAlex metadata and LLM categorizations. After all filtering and validation steps, 617 English-language journal papers with valid abstracts were retained for analysis. Manual inspection and cross-verification with metadata from Semantic Scholar and Crossref were conducted to correct erroneous metadata in OpenAlex, primarily targeting duplicated abstracts. %Further details on data collection and cleaning are provided \red{in supplementary materials (SM)~\ref{app:consistency}.} 

Once the abstracts and bibliographic information were consolidated, we employed GPT-o3-mini to identify papers relevant to the study objective. This was done using four sequential Yes/No questions, with one or two follow-ups for the first two questions \red{(see the supplementary data)}. The questions are used to select a concise set of abstracts such that the corpus fits the GPT-o3-mini context window of 200,000 tokens. Accordingly, 742 abstracts -- before the manual inspection of the abstracts -- were used to extract a representative list of categories for \textit{measures}, \textit{data types}, and \textit{research-question types}. Using the same LLM, each of 617 publications after the manual inspection of the abstracts was independently assigned to a category for these three dimensions. All the prompts can be found in the supplementary data.
%As detailed in \red{Section \ref{sec:consistency},} the coherence and consistency of these outputs -- both the generation and the assignment of the taxonomies -- were verified by LLM and human judges.  

These three paper properties were then used to construct a knowledge graph, where each paper is represented as a triplet of \{\texttt{measure, data type, research-question type}\}. In the graph, only these paper properties are represented as nodes, and other paper information is stored cumulatively as an edge property, if necessary. In doing so, we assume that each resulting triplet corresponds to a semantically coherent representation of a paper's content. This way, the structural relationships among measures, data types, and research-question types are expressed through this content-based co-occurrence knowledge graph.

\section{Results}
\begin{figure}[!thb]
    \centering
    \includegraphics[width=16cm]{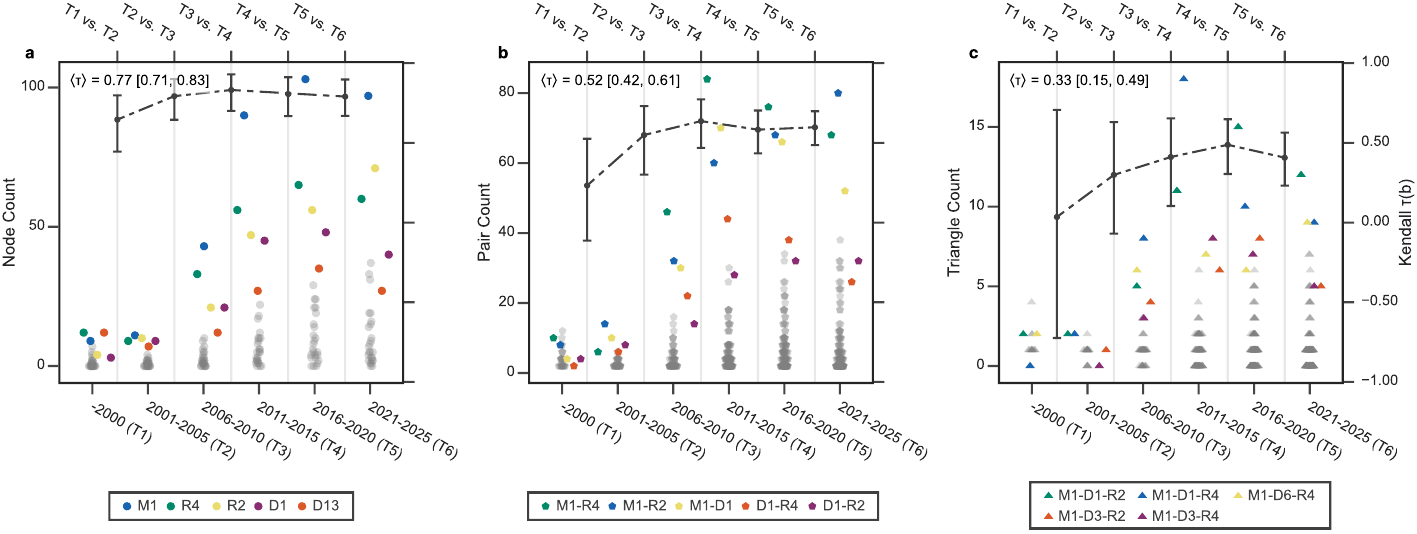}
    \caption{\textbf{Popular nodes, pairs, and triangles in the intergenerational wealth mobility literature across six periods.} 
    Each panel summarizes the distribution of components per period. Highlighted markers denote the five most frequent components, presented in the order of frequency. The black dashed line indicates Kendall's $\tau_B$ between two consecutive periods, computed over the common set of components. A 95\% bootstrap CI based on 1000 resamples is included to assess robustness given the small sample sizes per period. $\langle \tau \rangle$ denotes the mean observed coefficient across periods, with its 95\% bootstrap CI reported alongside.
    \textbf{a. Node occurrences (strength/2).} Between T3-T5, highlighted nodes dominate, with an increase in $\mathcal{M}_1$ and $\mathcal{R}_4$ at T3. At T6, the three gray nodes between $\mathcal{D}_1$ and $\mathcal{D}_{13}$ correspond to $\mathcal{D}_6, \mathcal{D}_2,$ and $\mathcal{M}_2$. Kendall's $\tau_B$ ranges between $0.64$ and $0.83$, indicating relatively high similarity between consecutive periods, and this similarity stays stable as reflected by the narrow bootstrap CIs (all above 0.5). 
    \textbf{b. Pairwise co-occurrences.} While $\mathcal{M}_1\text{-}\mathcal{R}_2$ and $\mathcal{M}_1\text{-}\mathcal{D}_1$ peak at T4 and decline thereafter, $\mathcal{M}_1\text{-}\mathcal{R}_2$ continues to increase and become dominant at T5. Kendall's $\tau_B$ is lowest between the first two periods ($0.23$) but stabilizes thereafter, reaching a maximum of $0.64$. Rank stability is low in the early comparison, and, the subsequent values remain above zero.
    \textbf{c. Three-way co-occurrences.} $\mathcal{M}_1\text{-}\mathcal{D}_1\text{-}\mathcal{R}_4$ peaks at T4 before falling below 10 observations by T6, allowing $\mathcal{M}_1\text{-}\mathcal{D}_1\text{-}\mathcal{R}_2$ to emerge as the dominant triangle from T5 onward. The first two periods exhibit minimal rank similarity ($\tau_B=0.04$). Similarity increases over time, albeit insignificantly, reaching a maximum of $\tau_B=0.5$, before slightly declining between the last comparison. Overall, stability remains low, particularly in the first three period-to-period transitions.
    }
    \label{fig:occurrences}
\end{figure}

\begin{figure}[!thb]
    \centering
    \includegraphics[width=12cm]{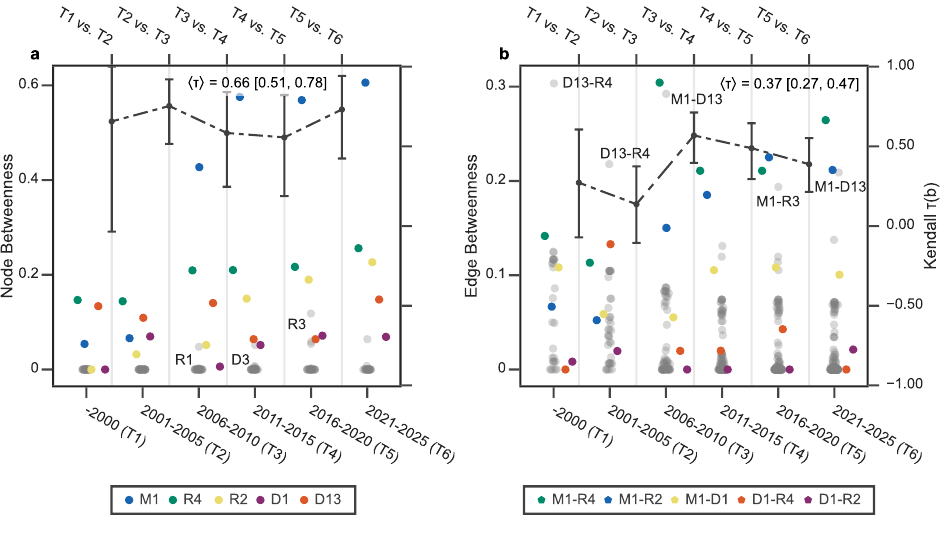}
    \caption{\textbf{Structurally central nodes and pairs in the intergenerational wealth mobility literature.} Each panel shows weighted betweenness centrality across periods. The same color scheme and Kendall's $\tau$ in the black dashed line as in Fig.~\ref{fig:occurrences} are used. Moreover, the same 95\% bootstrap CI and the mean observed coefficient across periods, $\langle \tau \rangle$, are calculated.
    \textbf{a. Weighted Node Betweenness.} $\mathcal{M}_1$ emerges as the most prominent bridging node from T3 onwards, suggesting that most papers have employed \textit{Regression-based Measures} ($\mathcal{M}_1$). Although highlighted, $\mathcal{D}_1$ exhibits lower betweenness than $\mathcal{R}_1$ (\textit{Measurement and Methodological Advances}) at T3, and both $\mathcal{D}_1$ and $\mathcal{D}_{13}$ fall below $\mathcal{R}_3$ (\textit{Policy, Institutional, and Geographic Impacts}) at T5. This indicates that high-frequency nodes do not necessarily occupy structurally diverse positions. Kendall's $\tau_B$ peaks at the comparison between T2 and T3 (0.75), despite the overlapping bootstrap CIs. All values remain above zero.
    \textbf{b. Weighted Edge betweenness.} The edge betweenness scores are generally lower than node betweenness, and the highlighted ones are no longer distinct from the gray ones. This suggests that pair-wise connectivity is less diversified, and that frequently occurring pairs do not always correspond to structurally central edges. At T1 and T2, $\mathcal{D}_{13}\text{-}\mathcal{R}_4$ (\textit{No dataset}-\textit{Intergenerational Wealth Mobility and Inheritance}) in gray attains the highest betweenness. At T3, $\mathcal{M}_1\text{-}\mathcal{D}_{13}$, records the second highest after $\mathcal{M}_1\text{-}\mathcal{R}_4$; at T5, $\mathcal{M}_1\text{-}\mathcal{R}_3$ follows the highlighted, $\mathcal{M}_1\text{-}\mathcal{R}_4$; and at T6, it follows $\mathcal{M}_1\text{-}\mathcal{R}_2$. $\mathcal{M}_1\text{-}\mathcal{R}_2$ (\textit{Regression-based Measures}-\textit{Empirical Estimates and Determinants}) increases steadily and becomes the highest at T5, whereas other periods are dominated by $\mathcal{M}_1\text{-}\mathcal{R}_4$ (\textit{Regression-based Measures}-\textit{Intergenerational Wealth Mobility and Inheritance}). The rank similarity between T3 and T4 appears to be highest at the median.
    } 
    \label{fig:betweenness}
\end{figure}

Fig.~\ref{fig:occurrences} summarizes the co-occurrence patterns among Measures ($\mathcal{M}$), Data types ($\mathcal{D}$), and Research-Question types ($\mathcal{R}$) across six consecutive periods. 
T3 (2006-2010) and onward, blue, green, and yellow of the highlighted nodes, pairs, and triangles -- representing the five most recurrent components of each type -- show their dominance across the panels. This dominance indicates that a small subset of methodological and conceptual combinations (e.g., $\mathcal{M}_1$, $\mathcal{R}_4$, and their pairwise or triadic combinations) increasingly shape the literature. Notably, the popularity of $\mathcal{M}_1$ (\textit{Regression-based Measures}), which begins to grow in T3, continues to increase, making it the only Measures element consistently present among the dominant pairs and triangles. 

The black-dashed line in each panel traces Kendall's $\tau_B$, that is, Kendall's $\tau$ that adjusts for ties, between consecutive periods. While rank similarity remains relatively high for nodes, it progressively decreases for pairs and triangles. This decreasing pattern is also stable across 1000 bootstrap resamples: the 95\% bootstrap confidence intervals (CIs) of the global mean for nodes are wider than those for pairs and triangles, although the difference between the pair and triangle CIs is less pronounced. Accordingly, the node popularity remains relatively stable over periods, whereas the popularity of their pairwise and triadic combinations varies more strongly. 

It is vital to note that this decrease in temporal stability is not entirely due to randomness or network size. We compared the observed counts to those in 1000 random models, where the number of nodes per period is preserved while assigning the three component labels at random. Part of the decrease in temporal stability (lower $\langle \tau \rangle$) in tandem with the increase in the component number (i.e., $\tau_{\text{node}} > \tau_{\text{pair}} > \tau_{\text{triangle}}$) can be attributed to network size, as the global mean appeared to decrease over the panels with the null models (See Fig.~\ref{fig:random_counts}). However, 
the node, pair, and triangle counts have their periodic Kendall's $\tau_B$ situated at a similar level, accounting for their Clopper-Pearson's 95\% CIs, and the differences in their global means appear smaller than the observed ones. Thus, the consistently and distinctively high similarity in rankings of node counts across periods seems to reflect genuine structural continuity in the literature rather than an artifact of size or randomness. More details on the random null models are given in \red{SM} \ref{app:robust}.

Together with the stability in node counts, the low $\tau_B$ values for pairs and triangles further support the fact that a small set of nodes form the dominant building blocks of higher-order structures. In other words, the dominant nodes, such as \textit{Regression-based Measures} ($\mathcal{M}_1$), \textit{Empirical Estimates and Determinants} ($\mathcal{R}_2$), and \textit{Intergenerational Wealth Mobility and Inheritance} ($\mathcal{R}_4$), combine with different tripartite components across periods, producing less stable rankings at the pair and triangle level. Lastly, the particularly low coefficients between T1 and T2 for pairs and triangles likely reflect the limited volume of pre-2000 literature, as also evidenced by the wide 95\% CIs from 1000 bootstrap samples.

The normalized weighted node and edge betweenness centralities in Fig.~\ref{fig:betweenness} show which nodes and edges serve as bridges in the network. In general, the betweenness scores for edges are remarkably lower than those for nodes, with the maximum score below 0.3. This suggests that the same edges may be shared by many triangles, and thus the components that distinguish different communities in the network are nodes, not edges. 

The generally lower betweenness scores at the edge level than the node level are also captured by the null models. The highest edge betweenness score lies around 0.047, and the highest node betweenness score lies around 0.08 -- both scores being significantly lower than the observed values (see Fig.~\ref{fig:random_betweenness}).  It can be inferred that, structurally, edge betweenness scores are lower than node betweenness scores, while the significantly higher node and edge betweenness may be due to the characteristics of the literature. 

The observed ranking of the betweenness between two consecutive periods seems to vary more than when considering the components' counts in Fig.~\ref{fig:occurrences}. The highest similarity is observed between T2 and T3 ($\tau_B=0.75$), with the node betweenness, and between T3 and T4 ($\tau_B=0.57$), with the edge betweenness. The generally lower Kendall's $\tau_B$ alludes to more changes in pairs' dynamics across the six periods than in the nodes. Note that the random null models also show generally lower Kendall's $\tau_B$ values at the edge level, yet to a lesser extent. 

Considering the node betweenness in Fig.~\ref{fig:betweenness}~\textbf{a}, $\mathcal{M}_1$ (\textit{Regression-based Measures}) appears to be most diversely connected, while $\mathcal{D}_1$ (\textit{Panel/Longitudinal Surveys}) -- albeit one of the most frequent nodes -- seems to be connected with a limited number of nodes of different partitions, given the node betweenness scores below 0.2 across all periods. Given the remarkably high count of $\mathcal{M}_1$ in T3 in Fig.~\ref{fig:occurrences}, it can be thought that \textit{Regression-based Measures} ($\mathcal{M}_1$) not only became more popular but also combined with different data types and research-question types, serving as a bridge for shortest paths between nodes. Moreover, across periods, only 3 to 7 nodes have the node betweenness scores above zero, which is contrary to the fact that more than $7$ nodes showed a node count above zero in Fig.~\ref{fig:occurrences}. These zero-scored nodes in Fig.~\ref{fig:betweenness} signify that they have appeared in only one triangle. In other words, they are not connected to other triangles with shared edges, which is further supported by the generally very low edge betweenness in Fig.~\ref{fig:betweenness}~\textbf{b}.

\begin{figure}
    \centering
    \includegraphics[width=16cm]{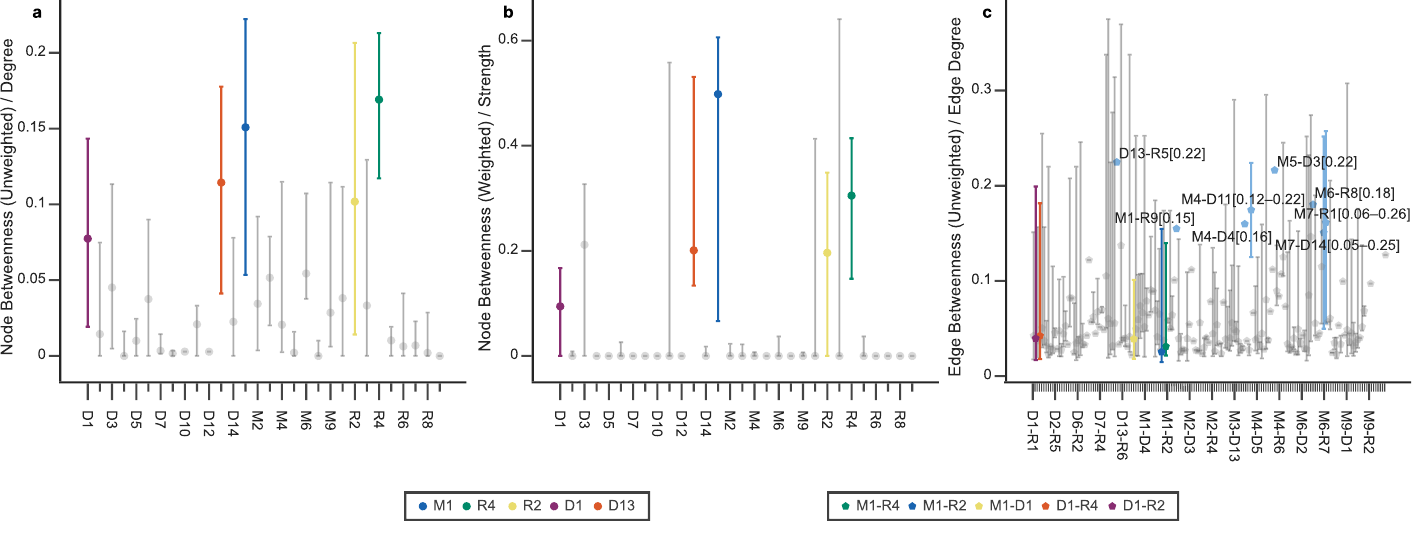}
    \caption{\textbf{Median betweenness-degree/count ratios.} Each point shows the period-median ratio for a node or pair. With the same color codes as above, the error bars are the range (min-max) of the ratio across six periods. Those that are not highlighted but have their ratio in the 95\textsuperscript{th} percentile are colored in light blue.  \textbf{a. Unweighted Node Capacitance (Eq.~\ref{eq:ratio_degree}).} Given that weights are ignored in calculating degrees, unweighted node betweenness is utilized for the ratio. The highlighted ones, which are the five most occurring nodes in the literature, not only have a high degree but also have a high unweighted node capacitance.
    \textbf{b. Weighted Node Capacitance (Eq.~\ref{eq:ratio_strength}).} Given that weights are considered in calculating strengths, weighted node betweenness is used for the ratio. Among the highlighted nodes, $\mathcal{D}_{13}$ is less frequently used in the literature than $\mathcal{D}_1$ but has a higher weighted node capacitance.
    \textbf{c. Unweighted Edge Capacitance (Eq.~\ref{eq:ratio_pair}).} Note that some edges only appear once in the network, hence no error bar. We consider the unweighted edge betweenness and edge degree. At the edge level, the five most popularly used elements have low median efficiency, unlike the other two panels. The highest capacitance on median is performed by $\mathcal{D}_{13}\text{-}\mathcal{R}_5$ (\textit{No dataset}--\textit{Demographic Differences in Mobility (Race, Gender, etc.)}). 
 }
    \label{fig:betweenness_ratios}
\end{figure}

\red{Lastly, combining the insights from the previous two figures, we turn to the capacitance measures in Fig.~\ref{fig:betweenness_ratios}. These capacitance measures highlight components that act as disproportionately strong bridges relative to their overall connectivity, thereby pointing to potentially underexplored combinations. Note that median values are shown in the figure, along with the range of ratios across six periods.} 

\red{The unweighted node capacitance (Eq.~\ref{eq:ratio_degree}) in Fig.~\ref{fig:betweenness_ratios}~\textbf{a} demonstrates that the five most frequently appearing nodes also exhibit high capacitance, even when disregarding their large number of partnering nodes. This suggests that, within the literature of intergenerational wealth mobility, \textit{Regression-based Measures}, research questions related to \textit{Intergenerational Wealth Mobility and Inheritance}, and \textit{Empirical Estimates and Determinants}, as well as data types such as \textit{Panel/Longitudinal Surveys} and \textit{No dataset}, occupy structurally central positions. When accounting for paper frequency, the weighted node capacitance (Eq.~\ref{eq:ratio_strength}) in Fig.~\ref{fig:betweenness_ratios}~\textbf{b}, on the other hand, reveals a striking contrast between these highlighted nodes and the rest, with the exception of $\mathcal{D}_3$, which displays a comparatively higher capacitance. Taken together, these results indicate that the most popular nodes act as key bridges between different triplets, in ways that cannot be explained solely by their degrees or strengths. This pattern points to a literature structured around a small set of dominant knowledge components (or paper properties) that recurrently connect otherwise distinct combinations of measures, data types, and research questions.}

\red{In terms of the unweighted edge capacitance (Eq.~\ref{eq:ratio_pair}) in Fig.~\ref{fig:betweenness_ratios}~\textbf{c}, the five most frequent pairs are not the ones with the highest capacitance. This means that the most popular pairs do not bridge different edges beyond their connectivity. Yet, it is noticeable that the edges with top 10\% capacitance (in light blue) are composed of two nodes that differ by at least 8 in their rankings of node degrees. The most considerable difference is by $\mathcal{M}_1 \text{-}\mathcal{R}_9$, where the first node has the highest degree, and the other has the lowest degree. The fact that no edges appear to have a distinctly high unweighted edge capacitance suggests that the network is well connected, which is likely the small corpus this paper focuses on. Within these edges, we further inspect the top two edges, i.e., $\mathcal{D}_{13}\text{-}\mathcal{R}_5$ and $\mathcal{M}_5\text{-}\mathcal{D}_3$. As they appear only once in a periodic network, we examine the impact of adding this edge to the periodic network before their appearance. Note that no changes will be made in the unweighted capacitance when the same triplet is added to the periodic network where the edge already exists.}

\red{For $\mathcal{D}_{13}\text{-}\mathcal{R}_5$, which appear as $\mathcal{M}_3\text{-}\mathcal{D}_{13}\text{-}\mathcal{R}_5$ in T3 (2006-2010), the addition of this triplet change the unweighted edge capacitance of 17 edges. Note that, besides $\mathcal{D}_{13}\text{-}\mathcal{R}_5$, the two other edges of this triplet had already appeared in the previous period. Most of these changes indicate that adding $\mathcal{M}_3\text{-}\mathcal{D}_{13}\text{-}\mathcal{R}_5$ reduce edge capacitance across the T2 (2001-2005) network, while four edges -- three of which are connected to $\mathcal{D}_{13}$ -- exhibit a slight increase in capacitance. In contrast, $\mathcal{M}_{5}\text{-}\mathcal{D}_3$ first appear as part of the triplet $\mathcal{M}_5\text{-}\mathcal{D}_{3}\text{-}\mathcal{R}_4$ in T2 (2001-2005), with all three edges making their first appearance in the network. In this case, adding the triplet to the T1 (-2000) network results in a reduction in capacitance across all 15 affected edges.}

\red{This investigation suggests that edges with high unweighted edge capacitance bridge the otherwise weakly connected areas of the network at the moment of their introduction, reducing many edges' capacitance. These two edges do not recur after their first appearance, suggesting that they are forgotten knowledge components in this field, despite their potential to connect different areas of the literature. In this sense, high-capacitance edges such as $\mathcal{M}_{5}\text{-}\mathcal{D}_3$ and $\mathcal{D}_{13}\text{-}\mathcal{R}_5$ could indicate \textit{forgotten combinations} of knowledge components that may warrant qualitative investigation, or highlight essential knowledge linkages that are structurally meaningful within the field. In other words, unweighted edge capacitance reveals hidden combinations of knowledge components that may later become recurrent pathways in the literature or define the knowledge space of the field.}

Two additional analyses in \red{SM}~\ref{app:results} can be applied. The first examines the relationship between degree and strength, as well as between triangle dispersion ratio (\red{SM}~\ref{app:triangle_dispersion_ratio}) and the number of unique triplets, thereby opening a discussion of the underlying structural configurations (Fig.~\ref{fig:relationships}). The second analysis incorporates the time-decayed appearance of nodes, pairs, and triangles. By allowing weights to decay over years in the case of absence, Fig.~\ref{fig:resurgence} can demonstrate when specific components resurge or progressively disappear over time.

\section{Discussion}
While the present analysis of the case study demonstrates the potential of the LLM-based content-level knowledge graph approach for revealing structural and temporal patterns in a research field, several limitations should be noted. 

First, the curated corpus of wealth mobility studies was intentionally narrow and domain-specific\red{, comprising 617 papers.} While the small dataset may limit us from generalizing the insights with the literature, this focused selection ensured conceptual consistency across nodes, pairs, and triangles. Moreover, we were able to focus on exploratory analyses in an environment free of any field- or topic-related factors. This way, the present study highlights the potential of our \red{LLM-empowered content-based knowledge graph for knowledge exploration.}

Second, the curation process relied on LLM-assisted classification of measures, data types, and research-question types based on the paper's abstracts rather than the full texts. Although it was a deliberate choice aimed at avoiding systematic bias toward open-access publications and our approach reached its internal consistency, utilizing the full texts would likely have increased the accuracy in the label generation and assignment. Nevertheless, these constraints do not detract from the central contribution of this work, demonstrating how content-based graph modeling can uncover the evolving architecture and methodological trajectories of a scientific field.

Lastly, the framework tested in this paper enforces each paper to be represented by one element for measures, data types, and research-question types. This design choice allows the resulting knowledge graph to remain structurally simple and analytically transparent, but it necessarily abstracts away some within-paper complexity. In particular, measures may be challenging to reduce to a single element, as some papers employ multiple measures or combine them. As a pragmatic solution, we extracted three categories of measures per paper, ordered by priority, and used only the highest-priority category when forming the triplet. Future work could extend the framework to accommodate multiple measures per paper without sacrificing interpretability.

Building on these findings, future research could extend the present framework in several directions. Expanding the corpus beyond wealth mobility studies to incorporate diverse research topics and domains would enable comparative analyses of methodological convergence and divergence across fields and of the influence of one field on another. Here, a more generalizable set of taxonomies can be designed. Moreover, while this paper's approach expressed each paper as a triplet, more than three components can be considered for other applications. 

Overall, we have demonstrated that turning citation networks inside-out, by focusing on the actual content of scientific papers rather than their citation relations, can give meaningful insights into the field of intergenerational wealth mobility and opens up opportunities for applications across other domains. \red{For instance, the capacitance measures have helped us identify key knowledge components in the literature, as well as the forgotten combinations of knowledge components that bridge otherwise weakly connected parts of the network.} All of which was possible thanks to the recent advancement in LLMs. More broadly, this approach suggests a shift in how we study and understand the structure of scientific knowledge, moving toward the organization of the ideas themselves.

\section{Materials and Methods} \label{sec:methods}

C1 of Fig. \ref{fig:overview} illustrates the methodological workflow of our case study. Here, we focus on papers' content to isolate patterns in how measures, data types, and research-question types co-occur. For each analysis, we construct a tripartite graph whose three disjoint node sets correspond to these categories. Thus, edges connect only across partitions and encode period-specific co-occurrence. To study temporal dynamics, we partition the corpus into size periods based on publication years: \texttt{pre-2000 (T1), 2001--2005 (T2), 2006--2010 (T3), 2011-2015 (T4), 2016--2020 (T5),} and \texttt{2021--2025 (T6)}. Note that the first period spans the years 1976-2000, given the small number of publications before 2000. Accordingly, the data structure is a multi-period multigraph: within each period, the network is simple (i.e., at most one weighted edge between any pair of nodes), but when all the periods are taken into account, the same pair of nodes may be connected by multiple parallel edges.

\subsection{Nodes}
Our corpus comprises 617 English-language papers in the Social Sciences. For each paper, we extract three categorical attributes from the abstract -- Measure ($\mathcal{M}$), Data type ($\mathcal{D}$), and Research-Question type ($\mathcal{R}$) -- which constitute the three disjoint node sets of the tripartite graph. All the categories are listed in Table \ref{tab:categories}. In total, 30 nodes are included in the graph with 8 $\mathcal{M}$, 13 $\mathcal{D}$, and 9 $\mathcal{R}$ nodes. One data type category, $\mathcal{D}_8$ (Rich List Data), is excluded because it does not appear in any of the 617 papers and would therefore be an isolated node with zero degree in every period. 

\begin{table}[!ht]
\tiny
\centering
\caption{\textbf{LLM-generated nodes and their codes.} Three categories -- Measures, Research-question types, and Data types -- are generated by GPT o3-mini based on abstracts. Although extracted, $\mathcal{D}_8$ (\textit{Rich List Data}) does not match any papers in the curated corpus of 617 papers.}
\begin{tabular}{clclcl}
\hline
  & Measures                                   &   & Research Question Types                                    &    & Data Type                            \\ \hline
M1 & Regression‐based Measures                  & R1 & Measurement and Methodological Advances                    & D1  & Panel/Longitudinal Surveys           \\
M2 & Rank‐based Measures                        & R2 & Empirical Estimates and Determinants                       & D2  & Administrative/Registry Data         \\
M3 & Transition Matrix / Probability   Measures & R3 & Policy, Institutional, and Geographic Impacts              & D3  & National Survey Data                 \\
M4 & Absolute Mobility Measures                 & R4 & Intergenerational Wealth Mobility and Inheritance          & D4  & Opportunity Atlas                    \\
M5 & Multigenerational Measures                 & R5 & Demographic Differences in Mobility (Race, Gender, etc.)   & D5  & Natural/Experimental Data            \\
M6 & Decomposition / Structural   Approaches    & R6 & Mobility and Non-Income Outcomes (Health, Wellbeing, etc.) & D6  & Linked Administrative Data           \\
M7 & Non‐parametric Approaches                  & R7 & Theoretical and Structural Models                          & D7  & International Panel Data             \\
M8 & Others\_Measure                            & R8 & Perceptions of Mobility and Attitudes                      & D8  & Rich List Data                       \\
  &                                            & R9 & Others\_RqType                                             & D9  & University/Institution Data          \\
  &                                            &   &                                                            & D10 & Pseudo-Panel/Household Budget Survey \\
  &                                            &   &                                                            & D11 & Archival/Historical Data             \\
  &                                            &   &                                                            & D12 & Big Data                             \\
  &                                            &   &                                                            & D13 & No dataset                           \\
  &                                            &   &                                                            & D14 & Others\_DataType                     \\ \hline
\end{tabular}
\label{tab:categories}
\end{table}

\subsection{Edges}
For each period $p$, we constructed a tripartite, undirected, weighted graph on the node sets $\mathcal{M}$, $\mathcal{D}$, and $\mathcal{R}$. We include exactly three edge types:
\texttt{[:CO\_MEASURE\_DATATYPE],[:CO\_DATATYPE\_RQTYPE] [:CO\_RQTYPE\_MEASURE]}
linking ($\mathcal{M, D}$), ($\mathcal{D, R}$), and ($\mathcal{R, M}$), respectively. Edges are undirected, and their weights capture period-specific document frequency: for eligible node pairs $u, v$,

\begin{equation}
    w_{uv}^{(p)} \;=\; \sum_{d\in \mathcal{C}_p} \mathbf{1}\!\left\{ u \in d \;\wedge\; v \in d\right\} , 
\end{equation}

where $\mathcal{C}_p$ is the corpus (set of papers) in period $p$ and $\mathbf{1\{\cdot\}}$ is the indicator function. Thus, multi-edges are collapsed and $w_{uv}^{p}\;=\;w_{vu}^{p}$ by construction; self-loops are excluded.

Consider two papers in the same period: the first contains ($\mathcal{M}_1, \mathcal{D}_1, \mathcal{R}_1$) and the second contains ($\mathcal{M}_1, \mathcal{D}_2, \mathcal{R}_1$). Then the following edges and weights are created:
\begin{align} \nonumber
&(\mathcal{M}_1, \mathcal{D}_1) \;:\; w=1, \; (\mathcal{D}_1, \mathcal{R}_1) \;:\; w=1, \; (\mathcal{R}_1, \mathcal{M}_1) \;:\; w=2, \\ 
&(\mathcal{M}_1, \mathcal{D}_2) \;:\; w=1, \; (\mathcal{D}_2, \mathcal{R}_1) \;:\; w=1.
\end{align}

\begin{figure}[!bht]
    \centering
    \includegraphics[width=9cm]{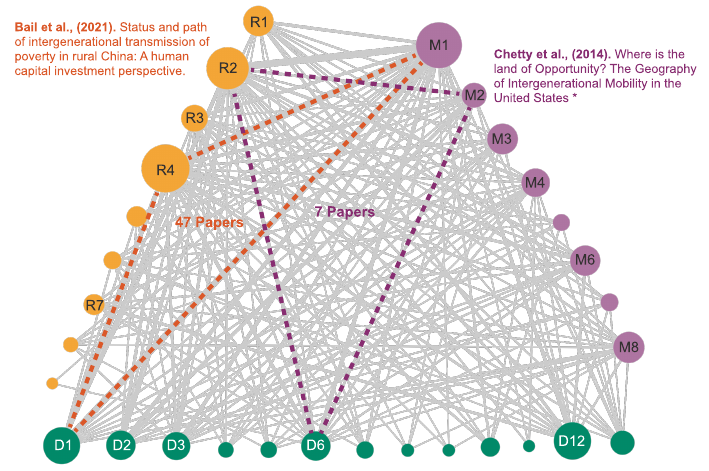}
    \caption{\textbf{Representing academic papers at the content-level summarizing all periods.} The knowledge graph is constructed by representing each of 617 papers as a triplet -- i.e., each paper consists of a measure, a data type, and a research question type. For example, Bail et al. (2021) paper is represented as  $\mathcal{M}_1$-$\mathcal{D}_1$-$\mathcal{R}_4$ and Chetty et al. (2014) paper as $\mathcal{M}_2$-$\mathcal{D}_6$-$\mathcal{R}_2$. Notice that although the used data structure is a multi-period multigraph, this visualization summarizes it across all six periods as a simple network.}
    \label{fig:graph_config}
\end{figure}

The resulting graph is presented in Fig.~\ref{fig:graph_config}. 

\subsection{Measures}
For each period $p$, we analyze the tripartite, undirected, weighted graph  $G_p \;=\; (V_p,E_p)$ on $\mathcal{M}$, $\mathcal{D}$, and $\mathcal{R}$, where $V_p$ comprises these three disjoint node sets and $E_p$ comprises the edges between these nodes. Let $w_{uv}^{p} \ge 0$ denote the period-specific edge weight between nodes $u$ and $v$, counting the frequency of the publications, and thus, $w_{uv}^{p} = 0$ if no edge exists.

The \textit{degree} is considered to see which properties are diversely connected. A high degree indicates that a node has a large number of neighboring nodes in a given period, denoting the \textit{breadth} of partnering. We note the degree of a node $x$ in period $p$ as $k_{x,p}$. To account for the paper frequency, the \textit{strength or weighted degree} will be calculated. It sums the edge weights incident to node $x$, illustrating the \textit{intensity} of the partnerships. We note the strength of the node $x$ in period $p$ as $s_{x,p}$. \red{Additionally, we consider \textit{edge degree} ($k_{e,p}$), which is defined as $k_{u,p}+k_{v,p}-2$ for an edge $e_{u,v}$ that connects nodes $u$ and $v$ in period $p$. A high edge degree indicates that the edge is connected to many nodes.}

Another centrality measure we use is betweenness centrality, both at the node and edge levels. While a high \textit{node betweenness centrality} tells us which node plays a central role in bridging otherwise distant parts of the network, a high \textit{edge betweenness centrality} gives information about the edge that plays the central role. Both of these are done by checking the number of shortest paths passing through, respectively, a node $x$ or an edge $e$ connecting nodes $u$ and $v$. We note the node betweenness as $B_{x,p}$ and the edge betweenness as $E_{e,p}$ for period $p$. Given that each edge is weighted by the occurrences in papers, we also calculate weighted betweenness at both levels by applying the inverse weight (i.e., $1/w_{uv}^{(p)}$). Thus, more frequently co-occurring properties are considered closer in distance. For the weighted, we note $B_{x,p}^{(w)}$ for node betweenness and $B_{e,p}^{(w)}$ for edge betweenness. Lastly, all the betweenness measures are normalized by the total number of distinct unordered
pairs, $\frac{2}{(n-1)(n-2)}$. We detail the definitions of these centrality measures in \red{SM}~\ref{app:measures}.

In this paper, we additionally compute three diagnostic ratios: (1) unweighted node betweenness divided by degree, (2) weighted node betweenness divided by strength, and (3) weighted edge betweenness divided by pair count \cite{EVERETT2016202bridging, wang2017cascading}. With these ratios, we aim to assess the extent to which nodes and edges exert structural influence relative to their local environment. These ratios highlight the cases where a node or edge is disproportionately central given its few neighbors, limited repeated interactions (i.e., paper frequency), or low co-occurrence frequency, as well as the converse patterns. As these ratios quantify the shortest-path flow a node or edge can support relative to its local structural capacity, we refer to them as \textit{capacitance} measures.

For a node $x$ in period $p$, the \textit{unweighted node capacitance} is defined as
\begin{equation}\label{eq:ratio_degree}
    R_{x,p} = \frac{B_{x,p}}{\tilde{k}_{x,p}},
\end{equation}
where $B_{x,p}$ denotes the unweighted betweenness of node $x$ in period $p$, and $\tilde{k}_x$ is its degree in the same period, normalized by the corresponding network size. A high unweighted node capacitance indicates that a node bridges a disproportionately large share of shortest paths relative to its number of direct connections. In other words, its effectiveness as a bridge is high. Note that both numerator and denominator are normalized, ensuring that for a star-shaped graph, the central node always has a ratio of 1, regardless of the network size.

The \textit{weighted node capacitance} is defined as
\begin{equation}\label{eq:ratio_strength}
    R_{x,p}^{(w)} =\frac{ B^{(w)}_{x,p}}{\tilde{s}_{x,p}},
\end{equation}
where $B^{(w)}_{x,p}$ is the weighted betweenness of a node $x$ in period $p$, and \red{$\tilde{s}_{x,p}$ is its strength (i.e., weighted degree) in the same period, normalized by the maximum strength in the corresponding periodic network.} As strengths are based on weights (i.e., paper frequency), the weighted betweenness is used here. A high weighted node capacitance implies that a node bridges a large portion of shortest paths relative to the total empirical weight of its connections or the total number of its popularity. As for the unweighted node capacitance, the central node of a star-shaped graph has a strength-wise ratio of 1. 

Lastly, for an edge $e = (u,v) \in E_p$ in period $p$, the \red{\textit{unweighted edge capacitance}} is defined as
\begin{equation}\label{eq:ratio_pair}
    R_{e,p}=\frac{B_{e,p}}{\tilde{k}_{e,p}},
\end{equation}
where $B_{e,p}$ denotes the unweighted betweenness of the edge $e$ between nodes $u$ and $v$ in period $p$, \red{and $\tilde{k}_{e,p}$ is the edge degree, normalized by $2(n-2)$. Pairs with high unweighted edge capacitance exert disproportionately strong bridging influence relative to the number of neighbors this edge has.}

\subsection{Connectivity of the Network}
Before turning to the results, we note that the knowledge graph exhibits no disconnected subgraphs in any of the six periods, nor in the aggregated network. All nodes are reachable through at least one path at every time slice. This guarantees that distance-based measures, most notably betweenness centrality and the ratios derived from it, are well-defined and comparable across periods with the normalization. 

At T1, 16 nodes with 32 edges; at T2, 18 nodes with 41 edges; at T3, 23 nodes with 64 edges; at T4, 27 nodes with 103 edges; at T5, 27 nodes with 120 edges; and at T6, 28 nodes with 123 edges are formed.

\subsection{Consistency and Reliability of LLM Outputs}\label{sec:consistency}

While LLMs offer an efficient means of categorizing a large number of papers, the generated categories may diverge from the intended classification. To assess the robustness of the assignments of LLM-generated categories to each paper, two complementary tests are conducted.

To assess the coherence of the LLM in category assignment, we compared two sets of LLM outputs. In the first set, categorization was performed using only the paper abstracts, while the second additionally included the paper titles in the input prompt. The two outputs largely overlapped in the assignment of measure categories, but diverged more strongly for data types and, more notably, for research-question types (Cohen's $\kappa_{\mathcal{M}}=0.85, \kappa_{\mathcal{D}}=0.82, \kappa_{\mathcal{R}}=0.70$). While this pattern again illustrates that the inherent fuzziness of the task, especially for conceptual classifications, can affect the consistency of the LLM's outputs, the substantial agreement level is reached in our case.

To verify the reliability of the LLM outputs for each paper, two annotators — one more experienced in wealth mobility studies and the other in economics — independently annotated 100 samples of the data, which were selected based on the number of citations. The agreement between the two annotators in terms of Cohen's $\kappa$ is 0.39 on average (Cohen's $\kappa_{\mathcal{M}}=0.38, \kappa_{\mathcal{D}}=0.49, \kappa_{\mathcal{R}}=0.31$). Annotator 1 was more aligned with the LLM's assignment for \textit{measures}, whereas more similarity was found with annotator 2 for \textit{data types} and \textit{research-question types}. We argue that this result does not undermine our experiment. Rather, it highlights the inherent difficulty of the labeling task and the challenge of maintaining consistency across a large corpus of papers, a setting in which LLMs can provide practical support.

\newpage

\bibliography{references}
\newpage

% \includepdf[pages=-, fitpaper=true]{SM_page.pdf}

\input{S0_Supplementary_Materials}

\end{document}

%% file: S0_Supplementary_Materials.tex
\setcounter{section}{0}
\setcounter{figure}{0}
\setcounter{table}{0}

\renewcommand{\thesection}{S\arabic{section}}
\renewcommand{\thefigure}{S\arabic{figure}}
\renewcommand{\thetable}{S\arabic{table}}

% Add the template cover for SM

\section*{Supplementary Text}
\addcontentsline{toc}{section}{Supplementary Text}

% \section{Consistency and reliability of LLM outputs} \label{app:consistency}
% \input{S1_Consistency}

\section{Definitions of Measures} \label{app:measures}
\input{S2_Measures}

\subsection{Additional Analytical Tools}\label{app:tools}
\input{S3_Additional_Tools}

\section{Unweighted Betweenness Centrality}\label{app:unweighted_btw}
\input{S4_Unweighted_Betweenness}

\section{Additional Analyses' Results}\label{app:results}
\input{S5_Additional_Analyses}

\section{Robustness Check}\label{app:robust}
\input{S6_Robustness_Check}

%% file: S2_Measures.tex
\paragraph{Unweighted Degree.} The \textit{degree}  of a node $x$ in period $p$ is given by
\begin{equation}
    k_{x,p} = |\{v \;\in\; V_{p}:w_{xv}^{(p)} >0\}|,
\end{equation}

i.e., the number of distinct neighbors connected to $x$ across the two opposite partitions (e.g., if $x$ is a $\mathcal{M}$, $x$ may be linked to different $\mathcal{D}$'s and $\mathcal{R}$'s. A high degree indicates that a node has a large number of neighboring nodes in a given period. When calculating the ratio of betweenness to degree, we normalize the degree by dividing it by the maximum degree in the network.

\paragraph{Strength or Weighted Degree.} The \textit{strength} is the sum of the weights of edges that $x$ is connected to
\begin{equation}
    s_{x,p} = \sum_{v \in V_p}w_{xv}^{(p)} \; ,
\end{equation}
which is also known as the weighted degree. In this paper, the weighted degree will be referred to as \textit{strength}, and the unweighted degree as \textit{degree}.

Strength increases when a node appears repeatedly with the same neighboring nodes (its partners), i.e., high co-occurrence frequency, even if the set of distinct partners is modest. Thus, the degree captures the \textit{breadth} of partnering, whereas the strength captures the \textit{intensity} of those partnerships.

\paragraph{Node Betweenness Centrality.}
The \textit{betweenness centrality} of a node $x$ in period $p$ is given by
\begin{equation}
    B_{x,p} \;=\; \sum_{\substack{s,t \in V_p \\ s \neq x \neq t}} \frac{\sigma_{st}(x)}{\sigma_{st}},
\end{equation}
where $\sigma_{st}$ denotes the total number of shortest paths between nodes $s$ and $t$, and $\sigma_{st}(x)$ is the number of those paths that pass through $x$. Nodes with higher $B_{x,p}$ values, therefore, play a more central role in bridging otherwise distant parts of the network. 

To account for edge weights, shorter paths were computed using the inverse of the co-occurrence frequency between nodes $u$ and $v$ (i.e., $1/w_{uv}^{(p)}$), such that more frequently co-occurring properties are considered closer in distance. For better comparison, the node betweenness centrality scores are normalized by the total number of distinct unordered pairs, $\frac{2}{(n-1)(n-2)}$ in this paper. Both weighted and unweighted node betweenness centralities are used in our analysis.

\paragraph{Edge Betweenness Centrality.}
The \textit{edge betweenness centrality} of an edge $e = (u,v) \in E_p$ in period $p$ is defined as
\begin{equation}
    B_{e,p} \;=\; \sum_{\substack{s,t \in V_p \\ s \neq t}} \frac{\sigma_{st}(e)}{\sigma_{st}} ,
\end{equation}
where $\sigma_{st}$ denotes the total number of shortest paths between node $s$ and $t$, and $\sigma_{st}(e)$ is the number of those paths that pass through edge $e$. Edges with higher $B_{e,p}$ values, therefore, act as key bridges between otherwise weakly connected regions of the graph, contributing to the overall structural cohesion of the network. Like the node betweenness centrality, the same inverse weight as in the node betweenness centrality  (i.e., $1/w_{uv}^{(p)}$) is used to calculate weighted edge betweenness scores, and the same normalization by $\frac{2}{(n-1)(n-2)}$ is applied in the analysis. Both weighted and unweighted edge betweenness centralities are used in our analysis.

\paragraph{Temporal Similarity of Component Ranks.} To quantify the similarity in counts, and betweenness centrality scores of nodes, pairs, and triangles between consecutive periods, we compute Kendall's rank correlation coefficient $\tau_B$, i.e., a Kendall's $\tau$ with adjustments for ties. For each component type, let $\mathbf{r}_p$ and $\mathbf{r}_{p+q}$ denote the rank vectors of component counts or scores in period $p$ and $p+1$, respectively. 

\begin{equation}
    \tau_B = \frac{n_c - n_d}{\sqrt{(n_0 - n_1)\,(n_0 - n_2)}},
\end{equation}
where $n_c$ and $n_d$ are the numbers of concordant and discordant pairs, $n_1$ and $n_2$ count ties in each ranking, and $n_0 = n(n-1)/2$ is the total number of pairs. The coefficient ranges from $-1$ (complete inversion) to $1$ (perfect agreement), with $\tau_B = 0$ indicating no association. We will refer to this as $\tau_B$ or simply, $\tau$ throughout the paper.

%% file: S3_Additional_Tools.tex
The following are the additional analytical tools that can be applied to larger corpora. 

\begin{figure}[!tbh]
    \centering
    \includegraphics[width=0.5\linewidth]{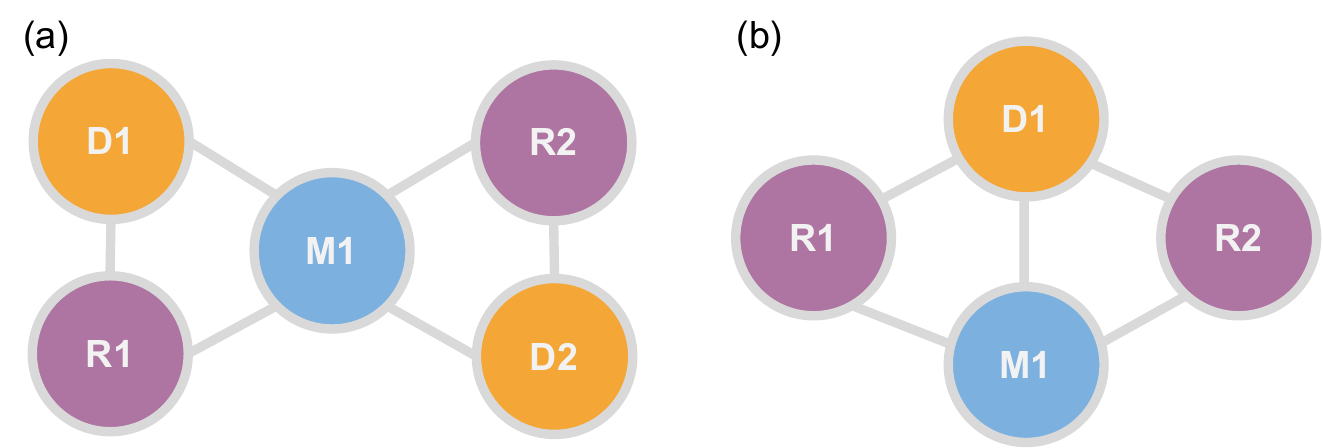}
    % \caption{Caption}
    \label{fig:triangle_dispersion_ratio}
\end{figure}

\paragraph{Triangle Dispersion Ratio.} \label{app:triangle_dispersion_ratio}
Another question that can be asked about the graph is which nodes comprise triangles using similar edges and distinct edges, which we will call \textit{triangle dispersion ratio}:
\begin{align}
    \eta_{x,p} = \frac{k_{x,p}}{2T_{x,p}},
\end{align}

where $T_{x,p}$ is the number of unique triplets $(m,d,r)$ in period $p$ such that $x$ is one of ${m,d,r}$, and $k_{x,p}$ is its degree, i.e., the number of distinct partners $x$ is connected to via the two opposite partitions in period $p$. As can be seen in the Figure above, if $x=\mathcal{M}_1$, two distinct triplets can either contribute to four edges around $x$ (panel (a), the triplets do not have a shared edge), or to three edges around $x$ (panel (b), the triplets have one shared edge in common). So, $2 T_{x,p}$ is an upper bound for the degree $k_{x,p}$. Thus, the \textit{triangle dispersion ratio} is bounded by $1$, with $\eta_{x,p}=1$ if and only if $x$ consists of only distinct triangles where no edges are shared.

The higher the $\eta$  is, the more connected the node $x$ is to different nodes, with more distinct triangles, like in panel (a) in the illustration above, whereas the lower the $\eta$, the more often the edges connected with the $x$ is reused to create different triads, like in panel (b).
In the case of high $T_{x,p}$, a node with a higher $\eta$ can be considered a mixer node, one that is frequently used and broadly combined across the two partitions. In contrast, a node with a lower $\eta$ is typically a node that is frequently used but mainly with a narrow, reused partner set. In the case of low $T_{x,p}$, a similar interpretation can be retrieved, but with the limitation that $x$ participates in a few realized three-way contexts.

\paragraph{Time-decayed Appearance Weight.}\label{para:time-decayed}
To capture the temporal persistence of key components -- nodes, pairs, and triangles, we define a time-decayed weight $\delta w_t$ that integrates past appearances with an exponential forgetting factor. For each component (node, pair, or triangle) and each year $t$, the decayed weight evolves recursively as
\begin{equation} \label{eq:decayweight_papercounts}
    \delta w_{t+1} = \begin{cases}
        \delta w_t\, e^{-\lambda} + c_{t+1}, & \text{if the component appears at year } t{+}1, \\
        \delta w_t\, e^{-\lambda}, &\text{otherwise,}
    \end{cases}
\end{equation}
where $\lambda>0$ is the decay rate and $c_{t+1}$ denotes the count of papers in which the component appears during year $t+1$. The initial condition is

\begin{equation}
    \delta w_{0} = \begin{cases}
        1, & \text{if the component first appears at year 0}, \\
        0, & \text{otherwise.}
    \end{cases}
\end{equation}

This formula ensures that recent appearances contribute more strongly to the weight, while earlier observations decay exponentially over time:
\begin{equation}
    \delta w_T = \sum_{t=0}^T c_t\,e^{-\lambda (T-t)}.
\end{equation}
The parameter $\lambda = \ln(2)/5$ corresponds to a half-life of five years, implying that the contribution of a past appearance is halved after five years of inactivity. Note that we consider all years between the minimum and maximum years in the dataset (i.e., 1976-2025), and that the decay is applied annually (i.e., $T-t = 1$). 

In the analysis, we examine two variants of time-decayed appearance weight: one where $c_{t+1}$ is the count of papers as (Eq.\ref{eq:decayweight_papercounts}) and another where $c_{t+1}=1$ if (re)appears at year $t+1$.

%% file: S4_Unweighted_Betweenness.tex
Here, we also present the unweighted betweenness centrality at both the node and edge levels.

\begin{figure}[t]
    \centering
    \includegraphics[width=12cm]{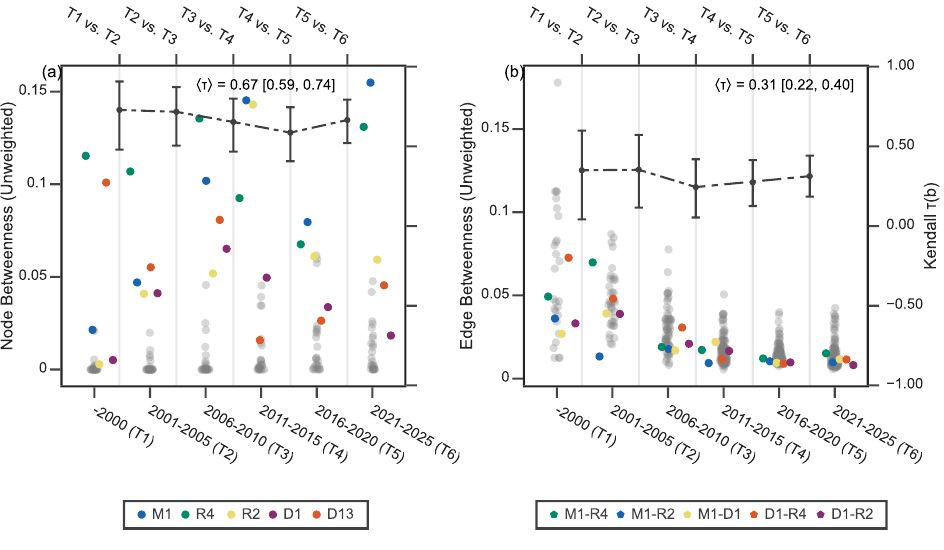}
    \caption{\red{\textbf{Unweighted Betweenness Centrality.} The same color scheme and Kendall's $\tau_B$ calculation as in previous figures in the main text are used in this figure. Overall, unlike in weighted betweenness centrality, unweighted betweenness centrality does not show a large difference between the node and edge levels on their y-axes. 
    \textbf{a. Unweighted Node Betweenness.} In the first three periods, $\mathcal{R}_4$ plays the most central node while in the last three periods, it is $\mathcal{M}_1$. Unlike the weighted betweenness in Fig.~\ref{fig:betweenness}, $\mathcal{R}_2$ also plays a crucial role in T4, and the gap between the most central node and the rest is not as big. The rankings of the betweenness appears to be strongly correlated across periods.
    \textbf{b. Unweighted Edge Betweenness.} It is more obvious than in the weighted edge betweenness in Fig.~\ref{fig:betweenness} that the highlighted ones are not necessary the ones with the high edge betweenness. Especially, in the last three periods, it appears that the popular pairs are not the ones that bridge areas of the network. The rankings of the betweenness appears to have a moderate similarity across periods.
    }}
    \label{fig:placeholder}
\end{figure}

%% file: S5_Additional_Analyses.tex
This section introduces the results of the additional analyses that can be applied to larger corpora. The conclusions drawn in the following subsections should not be generalized, given the limited size of our dataset. Nevertheless, we present these results to illustrate the kinds of analytical tools enabled by our content-based knowledge-graph approach.

\subsection{Degree, Strength, and Triangle Dispersion Relationships}

\begin{figure}[!thb]
    \centering
    \includegraphics[width=16cm]{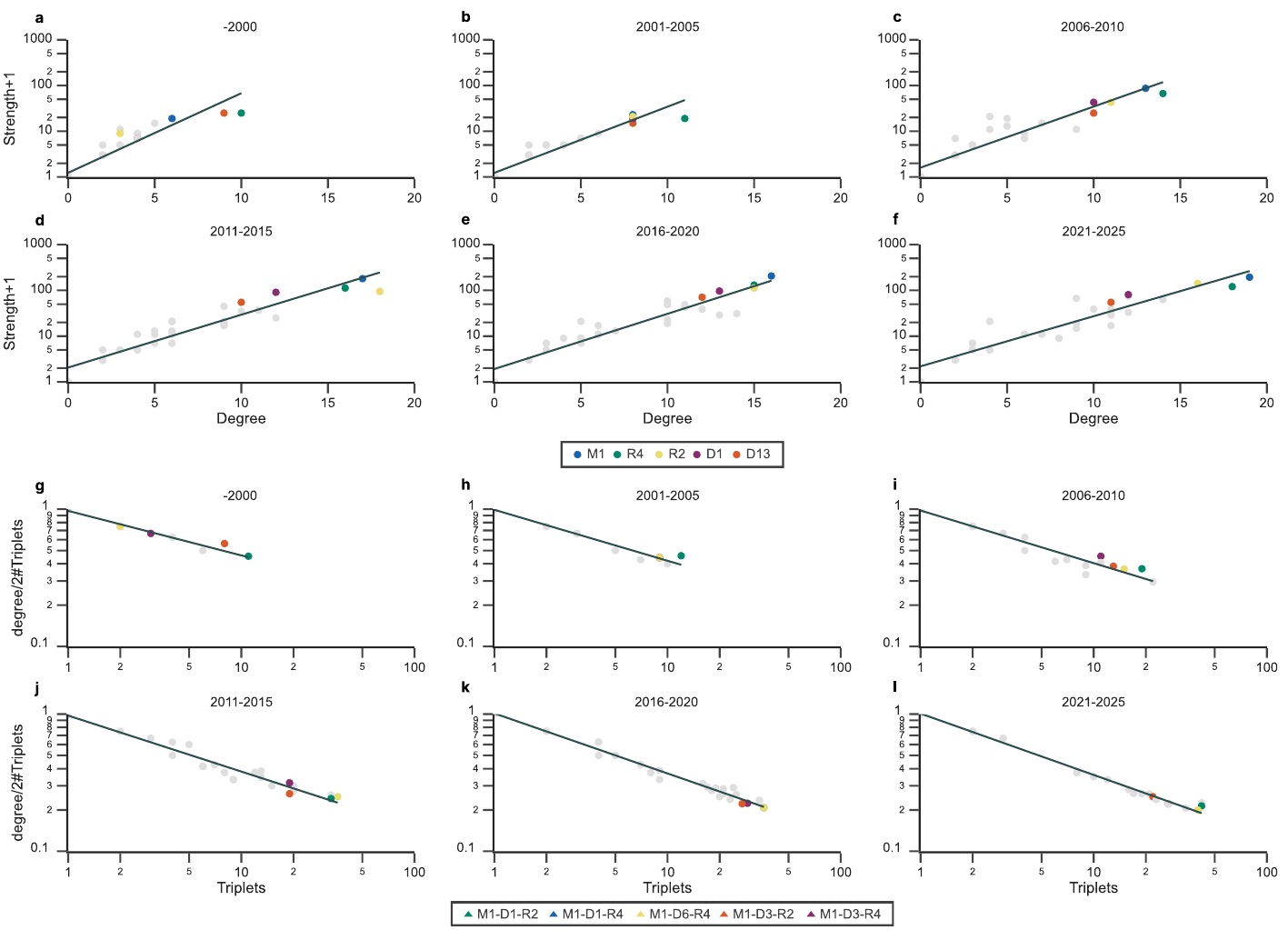}
    \caption{\textbf{Relationship between node degree and strength, and between triangle dispersion ratios and three-way co-occurrence patterns across periods.} Panels (a-f) show node degree versus strength $(+1)$, and panels (g-l) show Triangle Dispersion Ratio versus the number of distinct triangle co-occurrences for each period. Y-axes are logarithmically scaled to highlight proportional changes across orders of magnitude. The same coloring scheme as in Fig.~\ref{fig:occurrences} is applied.
    \textbf{a-f.} A clear positive association between degree and strength is observed: the higher a node's degree, the greater its strength, indicating that paper features connected to a wider range of other features also tend to co-occur more frequently in the literature. 
    \textbf{g-l.} A clear negative association emerges between the Triangle Dispersion Ratio and the number of unique triplets: nodes involved in many triplets tend to reuse the same edges across them, forming natural clusters of densely interconnected features.
    }
    \label{fig:relationships}
\end{figure}

Fig.~\ref{fig:relationships} summarizes the relationships between strength and degree, and between the triangle dispersion ratio and the number of unique triplets. It further supports that the higher the degree -- hence, the more diverse the node is connected to -- the more popular it is (i.e., the higher the strength). Moreover, as the number of unique triplets increases for a node, the number of edges shared to form triangles increases. Understood with Fig.~\ref{fig:betweenness}, this may hint that a few dominant nodes (\emph{e.g.}, $\mathcal{M}_1, \mathcal{R}_4, \mathcal{R}_2$) form dominant triangles with many shared edges. In other words, when studying wealth mobility and its methodology, most literature applies regression-based measures since the period of 2006-2010, on different types of data, often in order to answer research questions about intergenerational wealth mobility and inheritance.

\subsection{Resurgence}
\begin{figure}
    \centering
    \includegraphics[width=\linewidth]{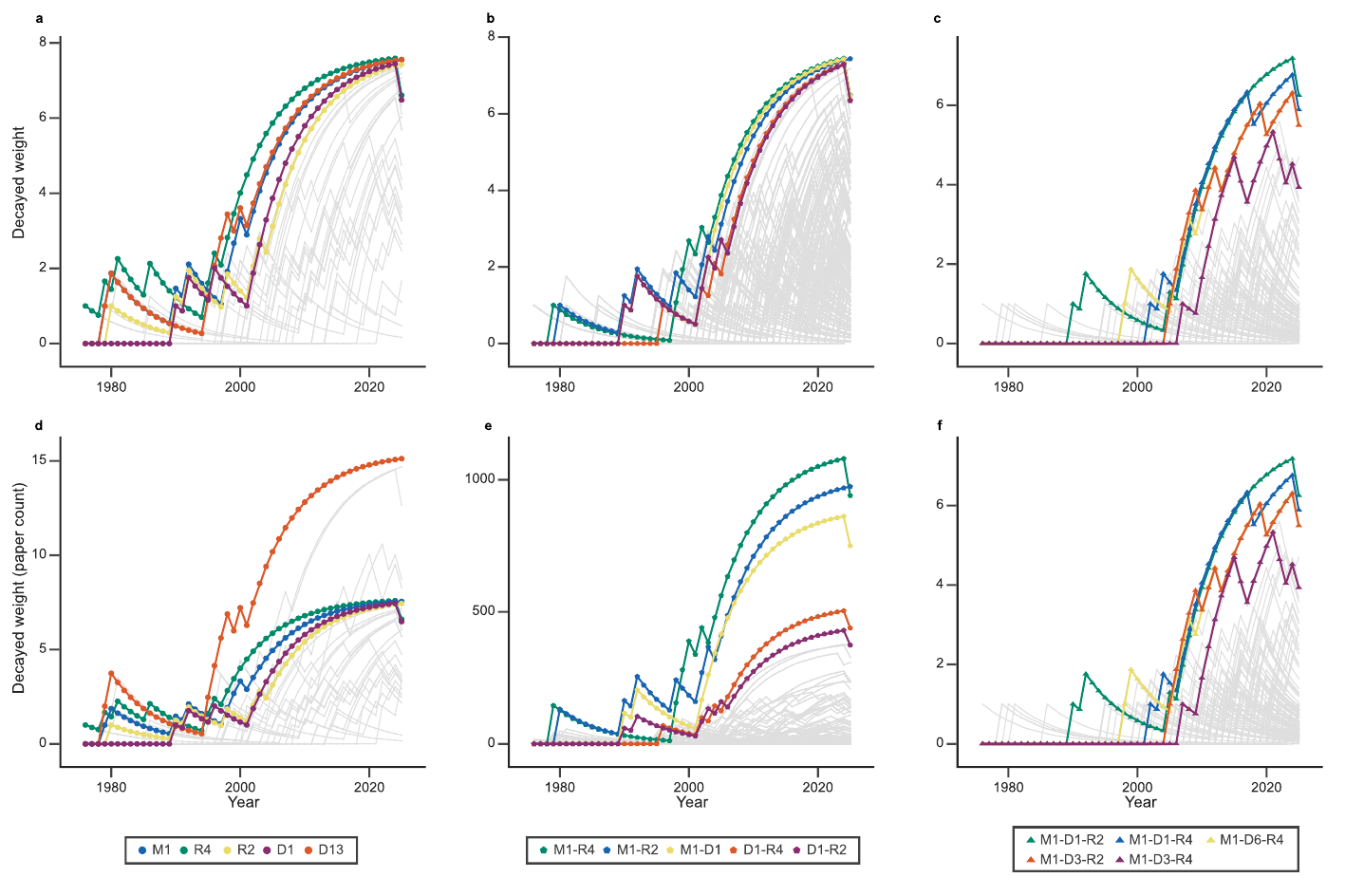}
    \caption{\textbf{Time-decayed appearance of nodes, pairs, and triangles with and without weighting by publication frequency over years.} The panels demonstrate the (re)appearance of nodes, pairs and triangles by applying the time-decayed appearance weights (see \ref{para:time-decayed}). Panels (a-c) only consider whether a key component appeared in a given year, while panels (d-f) additionally account for the number of papers containing that component. 
    % In both cases, weights decay exponentially across years with no subsequent observation after an appearance, following $\delta w_{t+1}=\delta w_t \cdot e^{-\lambda}$. If no appearance occurs, $\delta w_t =0$. In panels (a-c), each unique appearance of a key component in a year contributes one count ($\delta w_{t+1}=\delta w_t \cdot e^{-\lambda+1}$), whereas in panels (d-f), the count equals the number of papers featuring that component ($\delta w_{t+1}=\delta w_t \cdot e^{-\lambda+(\text{\#papers})}$). 
    \textbf{a\&d.} When considering only the net appearance of nodes, $\mathcal{R}_4$ remains the dominant node over the years, whereas $\mathcal{D}_{13}$ becomes dominant once paper frequencies are incorporated. In both panels, $\mathcal{R}_4$ makes a sharp decline after peaking in 1980, followed by a pronounced resurgence after 1994.
    \textbf{b\&e.} In both graphs, $\mathcal{M}_1\text{-}\mathcal{R}_4$ evolves into the dominant edge after a turning point in the year 1997, closely followed by $\mathcal{M}_1\text{-}\mathcal{D}_1$ and $\mathcal{M}_1\text{-}\mathcal{R}_2$. Given paper frequencies, the gap between these three widens.  $\mathcal{M}_1\text{-}\mathcal{R}_2$ shows more frequent appearances in later years than $\mathcal{M}_1\text{-}\mathcal{D}_1$. Distinct shifts are observed in 2001 for $\mathcal{M}_1\text{-}\mathcal{D}_1$, in 1997 for $\mathcal{M}_1\text{-}\mathcal{R}_4$, and in 1989 for $\mathcal{M}_1\text{-}\mathcal{R}_2$, with the latter showing the most abrupt evolution overall. 
    \textbf{c\&f.}The two graphs are identical, as no papers from the same year share the same triplets. $\mathcal{M}_1\text{-}\mathcal{D}_3\text{-}\mathcal{R}_4$
    appears latest among the highlighted triangles, whereas $\mathcal{M}_1\text{-}\mathcal{D}_1\text{-}\mathcal{R}_2$ emerges first in 1990. Compared to the nodes, where the gray traces closely follow the highlighted ones, edges and triangles (e,f) reveal stronger concentration, indicating that most papers focus on a limited set of dominant pairwise and triadic structures.}
    \label{fig:resurgence}
\end{figure}

Fig.~\ref{fig:resurgence} illustrates the resurgence patterns of key components based on two types of time-decayed appearance weights: one without considering paper frequencies (top) and one incorporating them (bottom). In both cases, it becomes evident that $\mathcal{M}_1$, previously identified as the most dominant node in terms of frequency and betweenness centrality, is not the most persistently reappearing feature throughout 1976-2025. Instead, $\mathcal{R}_4$, which first appeared in 1976, maintains a continuous presence from 1997 onward and emerges as the most persistent node by 2000. When the number of papers is taken into account, $\mathcal{D}_{13}$ exhibits a sharp decline between 1980 and 1994, suggesting a temporary shift toward other data sources, followed by a dramatic resurgence around 1998, reaching nearly twice the maximum weight of all other nodes and remaining dominant thereafter. These patterns highlight that the most popular component in terms of frequency is not necessarily the one that appears most consistently across time.
% Nodes show less temporal concistency than edges?

When examining the pairs, particularly those incorporating paper counts, a clear divergence emerges for $\mathcal{M}_1\text{-}\mathcal{R}_4$, $\mathcal{M}_1\text{-}\mathcal{R}_2$, and $\mathcal{M}_1\text{-}\mathcal{D}_1$, which separates markedly from the rest. This indicates that these three pairs are not only the most frequent across the six periods, but also persistent over time. Finally, the two panels for triangles are identical, as no year contains more than one paper representing the same triadic combination. In other words, within the wealth mobility literature, no two papers in the same year share exactly the same measure, data type, and research-question type.

%% file: S6_Robustness_Check.tex
\begin{figure}
    \centering
    \includegraphics[width=16cm]{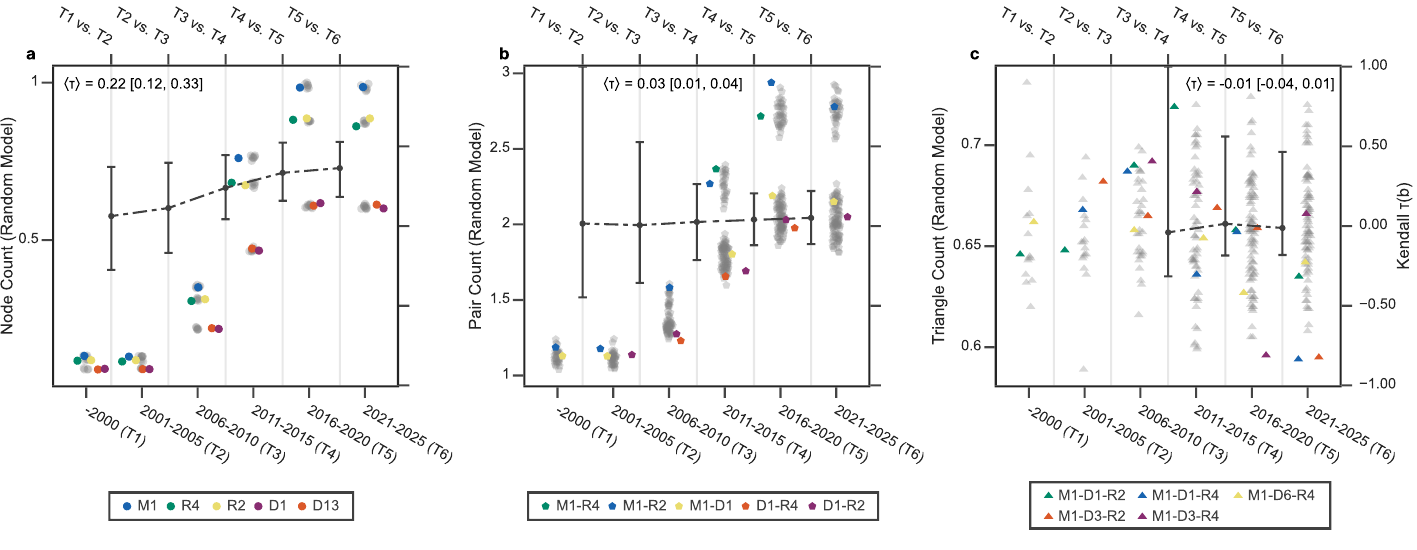}
    \caption{\textbf{Random null model for co-occurrences of components in the literature.} With the 1000 null samples, the node counts, pair counts, and triangle counts are plotted over six periods. At each period, the mean Kendall's $\tau_B$ over 1000 samples is plotted along with the Clopper-Pearson's 95\% CIs at each period. $\langle \tau \rangle$ is the global mean of the sampled Kendall's $\tau_B$, together with the bootstrap CI. Here, the bootstrap CI is used to achieve a coherence with Fig.~\ref{fig:occurrences}. Overall, across the panels, null distributions are significantly narrower and concentrated at substantially lower values than the observed ones. \textbf{a. Node counts.} In general, the period-wise mean Kendall's $\tau_B$'s and their corresponding CIs are lower than the observed ones in Fig.~\ref{fig:occurrences}~\textbf{a}. $\mathcal{M}_1$ remains as one of the leading nodes as of T3 as in the observed result, but together with other non-highlighted nodes. \textbf{b. Pair counts.} Still, $\mathcal{M}_1\text{-}\mathcal{R}_4$ and $\mathcal{M}_1\text{-}\mathcal{R}_2$ appeared in the leading group over periods. However, the green, $\mathcal{M}_1\text{-}\mathcal{R}_4$ is no longer one of the leading nodes in every period. A flat trend is observed in Monte Carlo procedure that continued until the Clopper-Pearson 95\% CI narrowed below 0.01 or until a maximum of 250,000 samples were reached. The Kendall's $\tau_B$ values are concentrated around zero over periods for, while the observed values were above zero in Fig.~\ref{fig:occurrences}~\textbf{b}. \textbf{c. Triangle counts.} The first two comparsions had undefined $\tau_B$ values due to insufficient overlap under the null model. Compared to the observed distribution in Fig.~\ref{fig:occurrences}~\textbf{c}, the null triangle counts are widely dispersed and the dominance of the highlighted triangles have disappeared.}
    \label{fig:random_counts}
\end{figure}

\begin{figure}
    \centering
    \includegraphics[width=12cm]{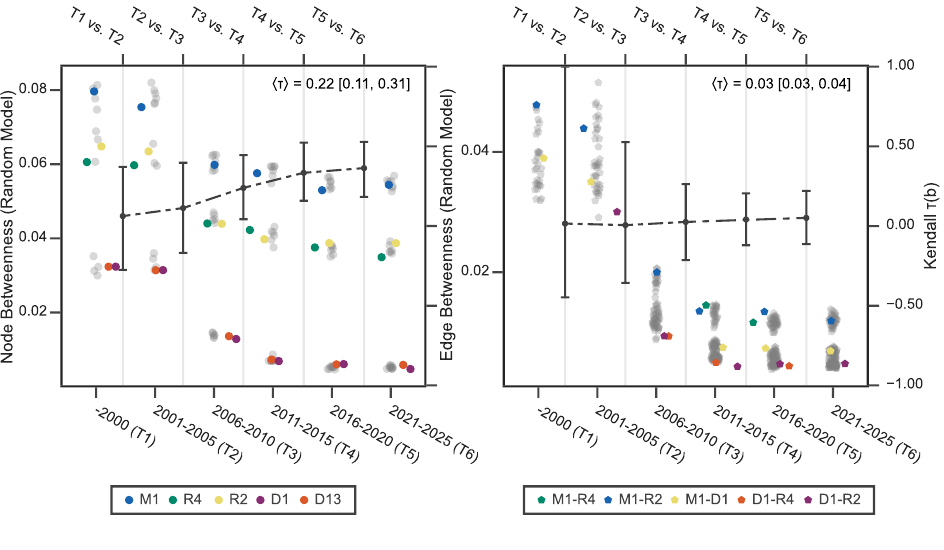}
    \caption{\textbf{Random null model for betweenness of components in the literature.} With the 1000 null samples, node and edge betweenness centrality scores are plotted over six periods. At each period, the mean Kendall's $\tau_B$ over 1000 samples is plotted, along with Clopper-Pearson 95\% CIs. $\langle \tau \rangle$ is the global mean of the sampled Kendall's $\tau_B$, together with the bootstrap CI. Here, the bootstrap CI is used to achieve coherence with Fig.~\ref{fig:betweenness}. Overall, at both node and edge levels, null distributions are concentrated at substantially lower values than the observed ones. \textbf{a. Node betweenness.} In general, Kendall's $\tau_B$ stays between 0 and 0.5, and the global mean is much lower than in the observed one in Fig.~\ref{fig:betweenness}.
    \textbf{b. Edge betweenness.} The first comparison shows a large variability across the null models, while it is narrower with the comparisons with later periods. Overall, the median $\tau_B$ value stays around zero across comparisons, while the observed ones ranged between 0.2 and 0.6.} 
    \label{fig:random_betweenness}
\end{figure}

Given the small sample size at each period, we generated a random null model to verify whether the observed patterns in the counts and the betweenness measures could be explained by network size. The randomization procedure preserves only the number of nodes per period while randomly assigning the three component labels. The core tripartite structure of the graph is maintained by ensuring that each paper receives one element of $\mathcal{M, D,} \text{ and } \mathcal{R}$. We then compared the observed Kendall's $\tau_B$ values across periods against the null node counts, pair counts, triangle counts, node betweenness, and edge betweenness. The number of null samples was determined adaptively using a sequential Monte Carlo procedure that continued until the Clopper-Pearson 95\% CI narrowed below 0.01 or until a maximum of 250,000 samples was reached.

At the node level, all measures across adjacent periods were significantly different from the null Kendall's $\tau_B$ values at 1000 samples. At the edge level, pair counts and edge betweenness scores (both weighted and unweighted) did not significantly differ from the null Kendall's $\tau_B$ for the first two comparisons and reached the Clopper-Pearson 95\% CI below 0.01 before reaching the 1000 samples. For the triangle counts, their first three comparisons are not significantly different from the observed Kendall's $\tau_B$, and Kendall's $\tau_B$ could not be computed. 

Fig.~\ref{fig:random_counts} demonstrates the results of the node, pair, and triangle counts with the 1000 null samples. The high global mean of Kendall's $\tau_B$ observed for node counts in Fig.~\ref{fig:occurrences} contrasts with the corresponding random model. Despite the wider 95\% CIs across periods, the global mean for pair counts ($0.03$) is substantially lower than the observed ($0.52$), with a bootstrap 95\% CI ($0.01\text{-}0.04$) which lies below the observed CI ($0.42\text{-}0.61$). For the triangle counts, the global means cannot be directly compared since Kendall's $\tau_B$ values for the first two null comparisons were undefined due to insufficient overlap under the null model. However, the remaining null comparisons cluster around zero, whereas the observed values are consistently positive. 

Overall, comparing the observed and null results suggests that part of the decrease in temporal stability (i.e,. lower $\langle \tau \rangle$) in tandem with the increase in the component number (i.e., $\tau_{\text{node}} > \tau_{\text{pair}} > \tau_{\text{triangle}}$) can be attributed to network size. Nevertheless, the consistently high similarity in node counts across periods appears to reflect genuine structural continuity in the literature rather than an artifact of size or randomness.